\documentclass[aps,pra,reprint,longbibliography]{revtex4-2}
\usepackage{amsmath,amsfonts,amssymb}
\usepackage{graphicx,bm}
\usepackage[breaklinks,colorlinks,urlcolor=blue,citecolor=blue,linkcolor=blue]{hyperref}

\newcommand{\vb}[1]{\mathbf{#1}} 

\newcommand{\ket}[1]{| #1 \rangle}

\newcommand{\expval}[1]{\langle #1 \rangle}
\usepackage{lipsum} 

\begin{document}
\title{Altermagnetism-induced Fulde-Ferrell-Larkin-Ovchinnikov superfluids and momentum-selective pairing channels in optical lattices}

\author{Qing-Yu Li$^{1,2}$}

\author{Zhen Zheng$^{1,2}$}\thanks{Corresponding author: zhenzhen@m.scnu.edu.cn}

\affiliation{
	$^1$Key Laboratory of Atomic and Subatomic Structure and Quantum Control (Ministry of Education), Guangdong Basic Research Center of Excellence for Structure and Fundamental Interactions of Matter, School of Physics, South China Normal University, Guangzhou 510006, China\\
	$^2$Guangdong Provincial Key Laboratory of Quantum Engineering and Quantum Materials, Guangdong-Hong Kong Joint Laboratory of Quantum Matter, Frontier Research Institute for Physics, South China Normal University, Guangzhou 510006, China
}

\begin{abstract}

We study the superfluid phases of a two-dimensional degenerate Fermi gas in an optical lattice subject to a $d_{xy}$-wave altermagnetic term that couples to the in-plane spin component. The single-particle Hamiltonian generates anisotropic Fermi surfaces for the two pseudo-spin species while maintaining zero net polarization. Using the Bogoliubov-de Gennes approach, we show that this anisotropy favors Cooper pairing at a finite center-of-mass momentum, producing a competition between Bardeen-Cooper-Schrieffer (BCS) and Fulde-Ferrell-Larkin-Ovchinnikov (FFLO) pairing and leading to FFLO superfluids. We further examine the effect of a Zeeman-induced spin polarization and find that triplet and singlet pairing channels coexist and compete, resulting in a momentum-selective pairing pattern. Remarkably, this triplet pairing emerges exclusively in the FFLO phase and is entirely absent in the BCS phase. The model is readily realizable in ultracold atoms with current experimental techniques, providing a feasible route for quantum emulation and for studying unconventional superfluid phases.

\end{abstract}
\maketitle

\section{Introduction}

Ultracold atomic systems have emerged as an ideal platform to explore quantum many-body physics \cite{Bloch2008JulRoMP,Safronova2018JunRoMP}, thanks to their high degree of controllability and flexible artificial manipulation.
With tunable interactions enabled by Feshbach resonances \cite{Koehler2006DecRoMP,Chin2010AprRoMP}, atoms can form Cooper pairs analogous to those in the Bardeen-Cooper-Schrieffer (BCS) state, leading to fermionic superfluids.
These systems are therefore well suited for developing theoretical insights into the pairing mechanism underlying conventional superconductors \cite{Giorgini2008OctRoMP}.
Beyond the conventional pairing scenarios, the engineering of synthetic gauge fields \cite{Dalibard2011NovRoMP,Goldman2014NovRoPiP,Halimeh2025JanNP,Zhang2018OctAP,Zhu2006DecPRL} in ultracold atoms can further pave the way for investigating unconventional quantum superfluid phases, which have attracted considerable recent interest \cite{Kamihori2021MayPRA,Inotani2021MayPRA,Vitali2022MayPRL,Huang2022JulPRB,Wu2023AprPRA,Liu2023NovPRL,Tu2024MarNJoP,Li2024MayPRA,Li2025MayPRA,Parida2025AugPRB,Stanescu2025NovPRA,Kobayashi2026FebPRB,Qiu2026MarTEPJD,Mosadeq2026MayPRB,Zheng2026JulNJoP,Beane2026AugPRA}.

In recent years, altermagnetism has emerged as a distinct type of magnetic order and has become a hot topic of current research \cite{Smejkal2022SepPRX,Smejkal2022DecPRX,Mazin2025JulAB}.
Fundamentally, altermagnetic order can be generated by spin-dependent hopping in ultracold atoms \cite{Das2024JunPRL,Zhu2025AprPRL,Liu2026FebPRB}, and its presence reorganizes the Fermi surfaces for different pseudo-spins \cite{Feiguin2009JulPRL}, making it possible to support unconventional pairing patterns rooted in specific symmetries \cite{Maeda2025AprPRB}.
The best-known theoretical proposal along this line is the Fulde-Ferrell-Larkin-Ovchinnikov (FFLO) state \cite{Fulde1964AugPR,Larkin1965SP}, in which Cooper pairs carry a finite center-of-mass (COM) momentum.
Several previous works have investigated FFLO pairing in related settings \cite{Zhang2024FebNC,Chakraborty2024AugPRB,Sim2025JulPRB,Yang2025DecPRB,Liu2026JanPRB},
but in most of these studies, the interaction is taken between bare pseudo-spins and therefore supports only singlet pairing \cite{Sigrist1991AprRoMP,Bergeret2005NovRoMP,Scalapino2012OctRoMP}.
The possibility that altermagnetism and interaction jointly generate diverse pairing channels has so far remained largely unexplored in the FFLO phase \cite{Fukaya2025AugJoPCM}.
Intuitively, triplet pairing can be introduced via a scattering interaction with an odd-parity pattern (e.g., $p$-wave \cite{Luciuk2016FebNP,Nagase2026JanPRL}), but it faces challenges in experimental engineering compared with the simple $s$-wave one.
While several works have proposed that modifying the multi-band band structure with complex orbital physics \cite{Zhu2023NovPRB,Ezawa2024OctPRB,Hong2025FebPRB,Pupim2025AprPRL,Yokoyama2026MayJotPSoJ,Mukasa2025JunJotPSoJ,Chatterjee2025AugPRB,Zhang2025SepPRB,Wang2026AprPRB,Li2026JulCPL,Yuan2026OctPRB} or optical fields \cite{Fu2026FebSP,Fu2026FebPRL,Yokoyama2025JulPRB} can offer alternative routes for hosting triplet pairing in altermagnetic systems,
it remains an open question whether a similar triplet pairing channel can emerge from a simple single-orbital model, and whether such a channel can coexist with or compete against the FFLO pairing.

To address these questions, in this paper, we study an altermagnetic system whose dressed pseudo-spins are quantized along $x$-axis.
The model system generates anisotropic Fermi surfaces and thereby strongly supports the FFLO state.
Unlike conventional multi-band models where triplet pairing typically relies on complex orbital physics or $p$-wave interactions, our system achieves the coexistence and competition of triplet and singlet pairing channels within a simple two-pseudo-spin model.
This is because the altermagnetic term projects the bare $s$-wave interaction onto the dressed pseudo-spin basis, effectively producing a momentum-selective pairing pattern.
Notably, this triplet pairing emerges exclusively in the FFLO phase, while it is absent in the BCS phase.
Since the model Hamiltonian is readily realizable in ultracold atoms with current experimental techniques, we expect that the results open an experimentally accessible route for quantum emulating and studying unconventional superfluid phases.

The paper is organized as follows.
In Sec. \ref{sec:model}, we first present the model Hamiltonian for the spin-balanced system.
Using the mean-field approach, we analyze the ground-state phase diagram and show the presence of the FFLO states.
In Sec. \ref{sec:momentum-selective-pairing}, we continue to study the effect of polarization on Cooper pairing. We find that polarization supports the coexistence and competition of the triplet and singlet pairing channels, and that the pairing mode becomes momentum-selective.
The Hamiltonians studied here can be realized in ultracold atoms, and we discuss the experimental realization in Sec. \ref{sec:experiment-realization}.
Finally, we summarize this work in Sec. \ref{sec:conclusion}.

\section{Model Hamiltonian}\label{sec:model}

We consider the degenerate Fermi gas loaded into a two-dimensional (2D) optical lattice. The Fermi gas consists of two internal atomic states, which we treat as pseudo-spins.
The Hamiltonian of the system is composed of three parts:
\begin{equation}
	H = H_0 + H_{\rm am} + H_{\rm int} \,. \label{eq:h-total-start}
\end{equation}
The first part describes the single-particle Hamiltonian:
\begin{equation}
	H_0 = -\sum_{\expval{i,j},\sigma} t \psi_{i\sigma}^\dag \psi_{j\sigma} - \mu \sum_{j,\sigma} \psi_{j\sigma}^\dag \psi_{j\sigma} \,,\label{eq:h-0-in-site-space}
\end{equation}
where $\psi_{\sigma}^\dag$ and $\psi_{\sigma}$ denote the creation and annihilation operators for atoms with spin $\sigma = \uparrow, \downarrow$, respectively.
$t$ is the magnitude of the nearest-neighbor hopping, and we use it as the energy unit throughout this paper.
The summation $\sum_{\expval{i,j}}$ runs over all pairs of nearest-neighbor sites.
$\mu$ is the chemical potential.
Here we study the spin-balanced case with equal $\mu$ for the two pseudo-spins.

The second part of the Hamiltonian \eqref{eq:h-total-start} describes the external potential that generates the $d_{xy}$-wave altermagnetic term (as will be shown latter) in the $\hat{S}_x$ direction. Its form is given by
\begin{equation}
	H_{\rm am} = \sum_{\vb{j}} \sum_{\eta,\eta'=\pm} (\eta\eta')\Omega_{\rm am} \psi_{\vb{j}\uparrow}^\dag\psi_{\vb{j}+\eta\vb{e}_x+\eta'\vb{e}_y,\downarrow} + H.c. \,,\label{eq:h-at-in-site-space}
\end{equation}
where $\Omega_{\rm am}$ denotes the strength of the potential.
$\vb{e}_{x}$ and $\vb{e}_{y}$ denote the unit vectors along the $x$ and $y$ axes, respectively, and $H.c.$ stands for the Hermitian conjugate.
Hamiltonian \eqref{eq:h-at-in-site-space} thus describes a spin-flip term between two next-nearest-neighbor sites, with the spin-flip strength acquiring a relative $\pi$ phase between hoppings along the $x\pm y$ directions.

The last part of the Hamiltonian \eqref{eq:h-total-start} describes the $s$-wave scattering interaction between opposite spins:
\begin{equation}
	H_{\rm int} = -U\sum_j \psi_{j\uparrow}^\dag \psi_{j\downarrow}^\dag \psi_{j\downarrow} \psi_{j\uparrow} \,,\label{eq:h-int-in-site-space}
\end{equation}
where the strength is denoted by $U$.
In this work, we consider the simple case of an attractive $s$-wave interaction ($U>0$), which is essential for supporting fermionic superfluids.

To extract the essential physics of the system, we map the Hamiltonian \eqref{eq:h-total-start} into momentum-$\vb{k}$ space. This yields
\begin{align}
	&H_0 = \sum_{\vb{k},\sigma}\xi_{\vb{k}}^{0}\psi_{\vb{k}\sigma}^\dag \psi_{\vb{k}\sigma} \,,\,
	H_{\rm am} = \sum_{\vb{k},\sigma}\xi_{\vb{k}}^{\rm am}\psi_{\vb{k}\uparrow}^\dag\psi_{\vb{k}\downarrow} + \text{H.c.} \,,\label{eq:h-single-particle-in-k-space}\\
	&H_{\rm int} = -U\sum_{\vb{k},\vb{k}',\vb{Q}} \psi_{\frac{\vb{Q}}{2}+\vb{k}\uparrow}^\dag \psi_{\frac{\vb{Q}}{2}-\vb{k}\downarrow}^\dag \psi_{\frac{\vb{Q}}{2}-\vb{k}'\downarrow} \psi_{\frac{\vb{Q}}{2}+\vb{k}'\uparrow} \,.\label{eq:h-int-in-k-space}
\end{align}
Here, $\xi_{\vb{k}}^{0} = -2t[\cos(k_x\lambda_L)+\cos(k_y\lambda_L)] - \mu$ and $\xi_{\vb{k}}^{\rm am} = 4\Omega_{\rm am}\sin(k_x\lambda_L)\sin(k_y\lambda_L)$, where $\lambda_L$ is the lattice constant.
One can identify $\xi_{\vb{k}}^{\rm am}$ as the $d_{xy}$-wave altermagnetic component \cite{Smejkal2022DecPRX}.
In Eq.\eqref{eq:h-int-in-k-space}, we have introduced $\vb{Q}$ as the COM momentum of the atomic pair during two-body scattering. One can verify that Eq.\eqref{eq:h-int-in-k-space} preserves momentum conservation.

\begin{figure}[t]
	\centering
	\includegraphics[width=0.48\textwidth]{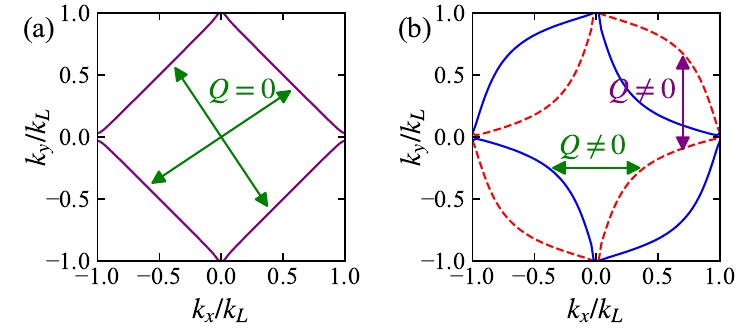}
	\caption{(a) Fermi surfaces without the altermagnetic term ($\Omega_{\rm am}=0$),
	which are isotropic and degenerate for the two pseudo-spins.
	(b) Fermi surfaces at $\Omega_{\rm am}=0.8t$. The blue solid lines and red dashed lines correspond to the Fermi surfaces of $\ket{+}$ and $\ket{-}$, respectively.
	The chemical potential is set to $\mu=0$.
	We set $\hbar=1$ in the whole paper and choose $k_L=\pi/\lambda_L$ as the momentum unit.}
	\label{fig:fermi_surfaces}
\end{figure}

Generally, when the attractive interaction $U$ exceeds a certain threshold, two atoms with opposite spins form Cooper pairs, giving rise to behavior reminiscent of superconducting order. Consequently, the fermionic system undergoes a transition from the normal gas phase to a superfluid phase.
To capture the many-body physics of the system, we first analyze the single-particle Hamiltonian $H_{\rm sp} = H_0 + H_{\rm am}$, as given in Eq.\eqref{eq:h-single-particle-in-k-space}.
The resulting Fermi surfaces in the momentum-$k$ space, obtained by diagonalizing $H_{\rm sp}$, are shown in Fig.\ref{fig:fermi_surfaces}.
When $\Omega_{\rm am} = 0$ as shown in panel (a), the Fermi surfaces are two-fold spin-degenerate. In this case, pairing is most efficient when the paired atoms carry opposite momenta, leading to a zero COM momentum $Q = 0$ in conventional systems.

However, when $\Omega_{\rm am} \neq 0$ that the altermagnetic term is present,  $H_{\rm sp}$ shares the eigenstates of the spin operator $\hat{S}_x$ if we choose that $\Omega_{\rm am}$ is real.
Hereafter we denote its eigenstates as $\ket{\pm}=(\ket{\uparrow}\pm\ket{\downarrow})/\sqrt{2}$, where $\ket{\sigma=\uparrow,\downarrow}$ denotes the bare spin states.
The system is then described by a two-band model. In this regime, the spin degeneracy is lifted for each band, resulting in split Fermi surfaces for the dressed states $\ket{\pm}$.
This is shown in Fig.\ref{fig:fermi_surfaces}(b).

It is worth noting that $H_{\rm sp}$ preserves the combined symmetry $C_{4z} \mathcal{T}$, where $C_{4z}$ represents a four-fold spatial rotation about the $z$-axis, and $\mathcal{T}$ denotes time-reversal symmetry \cite{Fu2026AugPRL}.
This symmetry gives rise to anisotropic features in the Fermi surfaces: the Fermi surface of the state $\ket{+}$ is elongated along the $k_x = -k_y$ direction, whereas the Fermi surface of the state $\ket{-}$ is elongated along the $k_x = k_y$ direction.
Such momentum-dependent anisotropy is a hallmark of altermagnetic order.
This symmetry also imposes that the net polarization in the $\ket{\pm}$ basis vanishes, giving rise to the characteristic altermagnetic behavior along the $\hat{S}_x$ direction.
Since Cooper pairing occurs between different bare pseudo-spins,
such anisotropic Fermi surfaces favor a nonzero COM momentum $Q \neq 0$ for inter-band pairing between different dressed states $\ket{\pm}$ as indicated in Fig.\ref{fig:fermi_surfaces}(b).
This observation motivates us to investigate the possibility of FFLO superfluids in this system.

We employ the mean-field Bogoliubov-de Gennes (BdG) approach. By introducing the superfluid order parameter $\Delta = -U \langle \psi_{\vb{Q}/2-\vb{k}\downarrow} \psi_{\vb{Q}/2+\vb{k}\uparrow} \rangle$.
In this work, the combined symmetry $C_{4z} \mathcal{T}$ plays a useful role in simplifying the calculation.
To proceed, we restrict our attention to four possible COM momenta \cite{Sim2025JulPRB}:
$\vb{Q}_1 = (Q, 0)$, $\vb{Q}_2 = (-Q, 0)$, $\vb{Q}_3 = (0, Q)$, $\vb{Q}_4 = (0, -Q)$,
which correspond to the four symmetry-equivalent directions related by the $C_{4z}$ rotation.
Then we can rewrite the interacting Hamiltonian \eqref{eq:h-int-in-k-space} in the following quadratic form.,
\begin{equation}
	H_{\rm int} = \sum_{\vb{k}}\sum_{\alpha=1}^4 (\Delta^\dag \psi_{\frac{\vb{Q}_\alpha}{2}-\vb{k},\downarrow} \psi_{\frac{\vb{Q}_\alpha}{2}+\vb{k},\uparrow}  + H.c.) + \frac{|\Delta|^2}{U} \,. \label{eq:h-pairing-4Q}
\end{equation}
The BdG Hamiltonian is then obtained by casting Eqs.\eqref{eq:h-single-particle-in-k-space} and \eqref{eq:h-int-in-k-space} into a matrix:
\begin{equation}
	H_{\rm BdG}(\vb{k}) = \begin{pmatrix}
		\hat{M}_{\vb{k}}^{0} \sigma_0 + \hat{M}_{\vb{k}}^{\rm am}\sigma_x & \hat{\Delta}_{\vb{k}}^\dag \\
		\hat{\Delta}_{\vb{k}} & -\hat{M}_{\vb{k}}^{0} \sigma_0 - \hat{M}_{\vb{k}}^{\rm am}\sigma_x
	\end{pmatrix} \,, \label{eq:h-bdg-in-k-space}
\end{equation}
where we have chosen the basis $\Psi = (\hat{\vb{u}}_{\vb{k}\uparrow}, \hat{\vb{u}}_{\vb{k}\downarrow}, \hat{\vb{u}}_{\vb{k}\uparrow}^\dag, \hat{\vb{u}}_{\vb{k}\downarrow}^\dag)^T$, with $\hat{\vb{u}}_\vb{k} = (\psi_{\vb{Q}_1/2+\vb{k},\sigma}, \psi_{\vb{Q}_2/2+\vb{k},\sigma}, \psi_{\vb{Q}_3/2+\vb{k},\sigma}, \psi_{\vb{Q}_4/2+\vb{k},\sigma})$.
The matrix symbols in Eq.\eqref{eq:h-bdg-in-k-space} are given by the diagonal matrix
$\hat{M}_{\vb{k}}^{\eta} = {\rm Diag}(\xi_{\frac{\vb{Q}_1}{2}+\vb{k}}^{\eta}, \xi_{\frac{\vb{Q}_2}{2}+\vb{k}}^{\eta}, \xi_{\frac{\vb{Q}_3}{2}+\vb{k}}^{\eta}, \xi_{\frac{\vb{Q}_4}{2}+\vb{k}}^{\eta})$ with $\eta$=0 or am,
and $\hat{\Delta}_{\vb{k}} = \Delta I_4 (i\sigma_y)$.
Here $I_n$ denotes the $n \times n$ identity matrix, and $\sigma_{x,y,z,0}$ are the Pauli matrices.

The thermodynamic potential at zero temperature is given by $\Omega_{\rm th} = \frac{1}{8}\sum_{\vb{k},\nu} E_{\vb{k}\nu}\Theta(-E_{\vb{k}\nu})+\mathcal{E}_0$, where $E_{\vb{k}\nu}$ denotes the energy of the $\nu$-th eigenstate of Hamiltonian \eqref{eq:h-bdg-in-k-space}, and the energy constant term $\mathcal{E}_0$ is defined as $\Omega_0 = \sum_{\vb{k}}\sum_{\alpha=1}^4 \xi_{\vb{Q}_\alpha/2-\vb{k}}^{0}/4+|\Delta|^2/U$.
By minimizing $\Omega_{\rm th}$ with respect to the order parameter $\Omega$ and the COM momentum $Q$,
we can determine the ground-state phase of the system.
Specifically, this is achieved by self-consistently solving the equations $\frac{\partial \Omega_{\rm th}}{\partial \Delta}=\frac{\partial \Omega_{\rm th}}{\partial Q}=0$.
The resulting ground state can be classified according to the values of $\Delta $and $\vb{Q}$.
If $\Delta\neq 0$, the system is in a superfluid phase: it corresponds to the FFLO superfluid phase when $\vb{Q} \neq 0$, and to the ordinary BCS superfluid phase when $\vb{Q} = 0$.
If $\Delta = 0$, the system reduces to the normal gas phase.

\begin{figure}[t]
	\centering
	\includegraphics[width=0.48\textwidth]{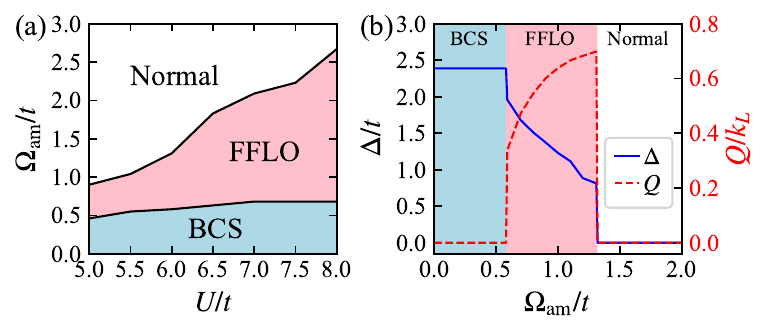}
	\caption{(a) Phase diagram with respect to $U$ and $\Omega_{\rm am}$.
	(b) The order parameters $\Delta$ (blue-solid line) and $Q$ (red-dashed line) at $U=6.0t$.
	We set $\mu = 1.0t$.}
	\label{fig:phase_diagram_without_zeeman}
\end{figure}

In Fig. \ref{fig:phase_diagram_without_zeeman}(a), we display the phase diagram in the $U$-$\Omega_{\rm am}$ plane, which provides a complete characterization of all three phases. The diagram is composed of three regions corresponding to the BCS superfluid, the FFLO superfluid, and the normal gas.
In particular, the FFLO phase occupies an intermediate window between the BCS and normal phases, and its extent is enlarged with the increase of $U$.
The changes of the order parameters $\Delta$ and $Q$ are displayed in Fig. \ref{fig:phase_diagram_without_zeeman}(b). One can see that, as $\Omega_{\rm am}$ increases, the order parameter $\Delta$ initially remains unchanged in the BCS phase, indicating that the pairing strength is essentially insensitive to the altermagnetic term as long as the system stays in the conventional BCS regime. However, after crossing a critical point at which the system transitions into the FFLO phase, $\Delta$ decreases monotonically with $\Omega_{\rm am}$, and eventually vanishes when $\Omega_{\rm am}$ exceeds a threshold value. This behavior reflects the fact that the altermagnetic term progressively distorts the Fermi surfaces and weakens the superfluid order. The disappearance of $\Delta$ at sufficiently large $\Omega_{\rm am}$ signals a transition into the normal gas phase, where the pairing instability is completely suppressed.

\begin{figure}[t]
	\centering
	\includegraphics[width=0.48\textwidth]{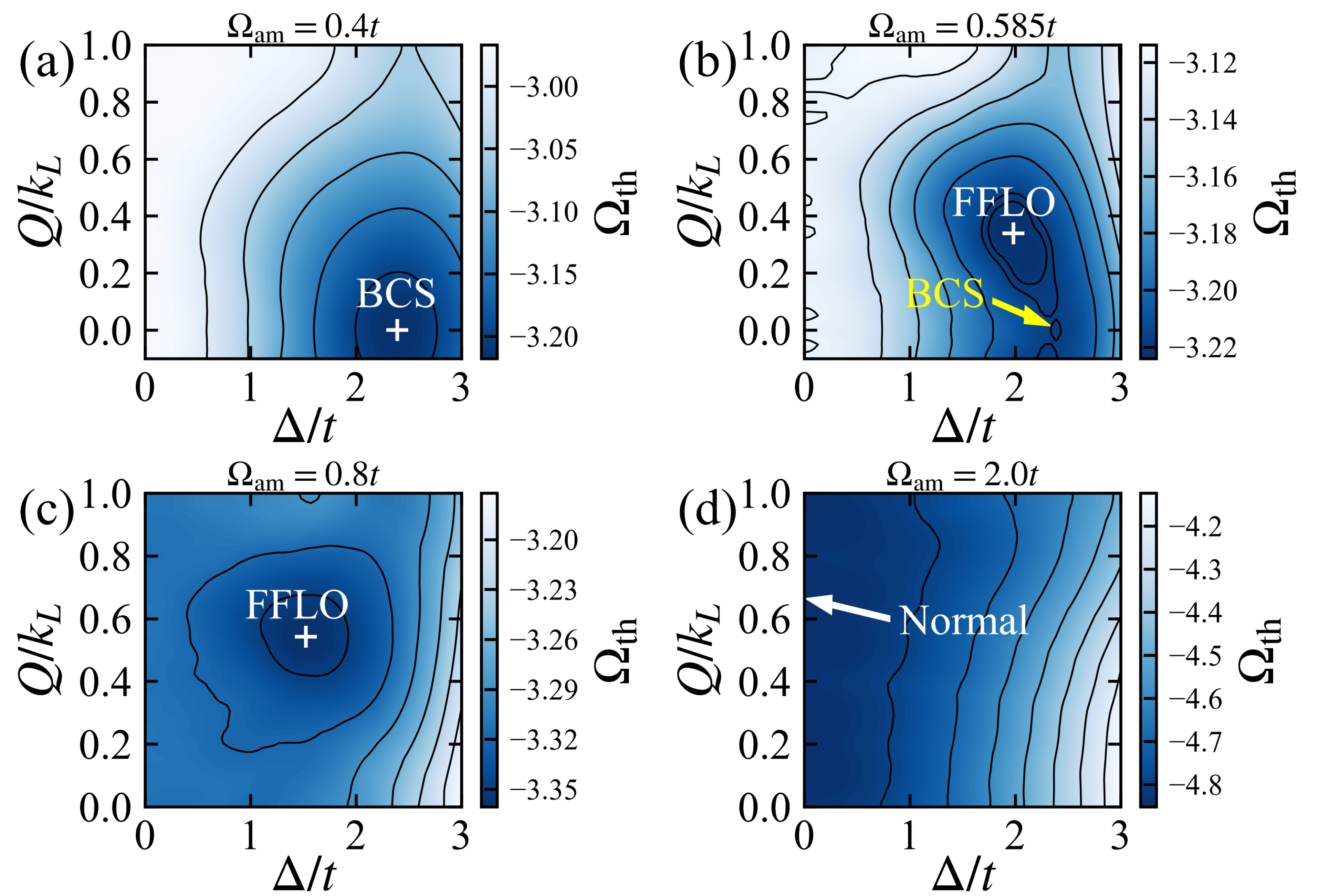}
	\caption{Contour plots of the thermodynamic potential $\Omega_{\rm th}$ as a function of the order parameters $\Delta$ and $Q$ for various values of $\Omega_{\rm am}$.
	The color represents the value of $\Omega_{\rm th}$, with darker regions indicating lower values.
	The plus symbols in (a)-(c), and white arrow in (d), mark the minima of $\Omega_{\rm th}$, which correspond to the ground states.
	The coexistence of two local minima corresponding to the BCS (indicated by the yellow arrow) and FFLO (white plus symbol) states are shown in (b).
	Other parameters are the same as in Fig.\ref{fig:phase_diagram_without_zeeman}(b).}
	\label{fig:contour_map}
\end{figure}

The transition between the BCS and FFLO phases can be further examined through contour plots of the thermodynamic potential $\Omega_{\rm th}$ for various parameter setups. In Fig. \ref{fig:contour_map}, we show the numerical results in the order-parameter plane of $\Delta$ versus $Q$.
As $\Omega_{\rm am}$ increases from zero, two local minima emerge and coexist, corresponding respectively to the BCS and FFLO phases.
This indicates that a competition exists between the two stable solutions, and that, after crossing a critical point, the phase transition takes place.
The coexistence of these two minima near the transition point suggests that the BCS-FFLO transition is of first order, since the two phases are separated by a finite barrier in the thermodynamic potential.
This is consistent with the discontinuous changes of $\Delta$ as shown in Fig.\ref{fig:phase_diagram_without_zeeman}(b).

\section{Momentum-selective pairing}\label{sec:momentum-selective-pairing}

In the preceding discussions, we have focused on the spin-balanced system with zero net polarization. We now turn to the effect of polarization on the FFLO phase induced by altermagnetism, which constitutes a natural and physically relevant extension of our analysis.

In ultracold atomic systems, polarization is typically introduced through a Zeeman field that acts oppositely on the two pseudo-spin components:
\begin{equation}
	H_z = \sum_{j} h_z(\psi_{j\uparrow}^\dag \psi_{j\uparrow} - \psi_{j\downarrow}^\dag \psi_{j\downarrow}) \,,
	\label{eq:h-zeeman-in-site-space}
\end{equation}
which is added to the total Hamiltonian \eqref{eq:h-total-start}.
Here, $h_z$ denotes the magnitude of the polarization, which controls the degree of spin imbalance between the two pseudo-spin species.
Accordingly, the BdG Hamiltonian \eqref{eq:h-bdg-in-k-space} is modified as
\begin{equation}
	H_{\rm BdG}'(\vb{k}) = H_{\rm BdG}(\vb{k}) + h_z \tau_z I_4\sigma_z \,,
\end{equation}
where $\tau_z$ denotes the Pauli matrix defined in the particle-hole basis.
We remark that the presence of this term \eqref{eq:h-zeeman-in-site-space} explicitly breaks the spin degeneracy between the bare pseudo-spins $\ket{\uparrow\downarrow}$,
whereas the net polarization remains zero in terms of the dressed pseudo-spins $\ket{\pm}$.

In the presence of the Hamiltonian \eqref{eq:h-zeeman-in-site-space}, the eigenstates of the single-particle Hamiltonian $H_{\rm sp}' = H_0 + H_{\rm am} + H_z$ in momentum space are written as
\begin{equation}
	\begin{cases}
		\psi_{\vb{k},+} = \cos\frac{\theta_{\vb{k}}}{2}\psi_{\vb{k}\uparrow}
		+ \sin\frac{\theta_{\vb{k}}}{2}\psi_{\vb{k}\downarrow} \\[6pt]
		\psi_{\vb{k},-} = -\sin\frac{\theta_{\vb{k}}}{2}\psi_{\vb{k}\uparrow}
		+ \cos\frac{\theta_{\vb{k}}}{2}\psi_{\vb{k}\downarrow}
	\end{cases} \,,\label{eq:operator-transfer}
\end{equation}
where
$\sin\frac{\theta_{\vb{k}}}{2} = \sqrt{\frac{1}{2}[ 1 - \frac{\hat{d}_z(\vb{k})}{|\hat{d}(\vb{k})|} ]}$,
$\cos\frac{\theta_{\vb{k}}}{2} = \sqrt{\frac{1}{2}[ 1 + \frac{\hat{d}_z(\vb{k})}{|\hat{d}(\vb{k})|} ]}$
with $\hat{d}_z(\vb{k}) = h_z$, $\hat{d}_x(\vb{k}) = \xi_{\vb{k}}^{\rm am}$, and $|\hat{d}(\vb{k})| = \sqrt{|\hat{d}_z(\vb{k})|^2 + |\hat{d}_x(\vb{k})|^2}$. Here, the mixing angle $\theta_{\vb{k}}$ characterizes the relative weight of the two pseudo-spin components in each eigenstate, and is determined by the competition between the Zeeman field $h_z$ and the altermagnetic term $\xi_{\vb{k}}^{\rm am}$.

Under the transformation \eqref{eq:operator-transfer}, the energy of the single-particle Hamiltonian $H_{\rm sp}'$ is given by $\xi_{\vb{k}}^0 \pm |\hat{d}(\vb{k})|$, and the pairing Hamiltonian from Eq. \eqref{eq:h-bdg-in-k-space} is written as
\begin{align}
	H_{\rm int} &= \sum_{\vb{k},\alpha} \Big\{
		\Delta \Big[ B_1(\vb{k}, \vb{Q}_\alpha) \sum_{\nu=\pm}\psi_{\frac{\vb{Q}_\alpha}{2}-\vb{k},\nu} \psi_{\frac{\vb{Q}_\alpha}{2}+\vb{k},\nu} \notag\\
		& + B_2(\vb{k}, \vb{Q}_\alpha) \psi_{\frac{\vb{Q}_\alpha}{2}-\vb{k},-} \psi_{\frac{\vb{Q}_\alpha}{2}+\vb{k},+} \Big] + H.c.
	\Big\} + \frac{|\Delta|^2}{U} \,, \label{eq:h-int-4Q-in-zeeman-case}
\end{align}
where the explicit forms of the $\vb{k}$-dependent modes $B_{1,2}(\vb{k}, \vb{Q})$ are given in the Appendix \ref{sec:pairing-modes}.

As is evident from Eq.\eqref{eq:h-int-4Q-in-zeeman-case}, after projecting the bare $s$-wave interaction onto the dressed basis, the mixing of pseudo-spin states induced by the altermagnetic term now acquires two pairing channels. The first describes the triplet pairing within the same eigenstate, characterized by an odd-parity mode $B_1(\vb{k})$ (hereafter we omit the variable $\vb{Q}$ in $B_{1,2}$ for brevity unless ambiguity arises). This pairing channel corresponds to the intra-band pairing illustrated by the purple arrows in Fig.\ref{fig:fermi_surfaces}(b). The second describes the singlet pairing between different eigenstates, characterized by an even-parity mode $B_2(\vb{k})$. This channel corresponds to the inter-band pairing illustrated by the green arrows in Fig.\ref{fig:fermi_surfaces}(b).
We demonstrate that $B_1(\vb{k})$ exhibits a nonzero distribution only when $\vb{Q} \neq 0$ and $h_z \neq 0$ (see Appendix \ref{sec:pairing-modes}). It indicates that the key to the emergence of triplet pairing is the polarization term $h_z$ together with the finite COM momentum.
Therefore the triplet pairing emerges exclusively in the FFLO phase, while is entirely absent in the BCS phase.

\begin{figure}[t]
	\centering
	\includegraphics[width=0.48\textwidth]{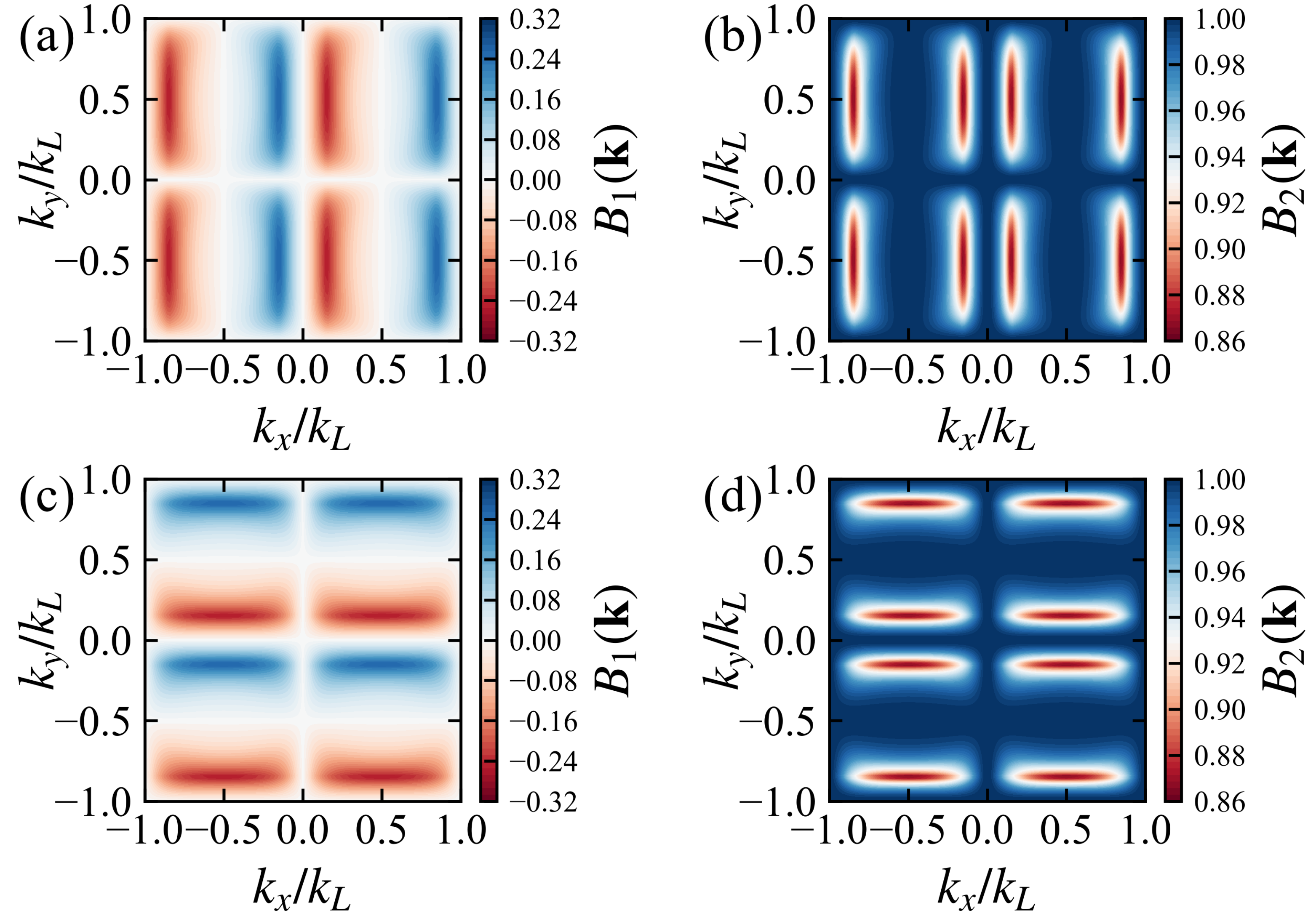}
	\caption{Distribution of pairing modes $B_{1,2}(\vb{k})$ in the $k_x$-$k_y$ plane.
	The COM momentum is chosen as $\vb{Q}=(Q,0)$ for (a)-(b) and $(0,Q)$ for (c)-(d), with $Q=0.3k_L$.
	Other parameters are $\Omega_{\rm am}=0.5t$ and $h_z=0.2t$.}
	\label{fig:pairing_func}
\end{figure}

In Fig.\ref{fig:pairing_func}, we display the distributions of the pairing modes $B_{1,2}(\vb{k})$ in the first Brillouin zone (BZ).
The odd-parity mode $B_1(\vb{k})$, associated with the intra-band triplet pairing, exhibits a pronounced sign-changing pattern, as shown in Fig.\ref{fig:pairing_func}(a) and (c).
In contrast, the even-parity mode $B_2(\vb{k})$, which governs the inter-band singlet pairing, remains strictly positive across the entire BZ, as seen in Fig.\ref{fig:pairing_func}(b) and (d).
One can see that this $k$-space structure reveals a direct competition between the singlet and triplet pairing channels: in the region where the triplet pairing is enhanced, the singlet pairing is suppressed.
Therefore the effective pairing strength becomes highly dependent on the momentum $\vb{k}$.
For $\vb{Q}$ along the $x$-axis, i.e. Fig.\ref{fig:pairing_func}(a) and (b), the weight of both pairing modes is concentrated along the $k_y$ axis.
When $\vb{Q}$ is rotated to the $y$-axis, i.e. Fig.\ref{fig:pairing_func}(c) and (d), the distribution rotates accordingly and is elongated along the $k_x$ axis.
Such momentum selectivity is a hallmark of the interplay between FFLO pairing and the altermagnetic term.

We remark that the triplet pairing component identified in this work could serve as a sensitive probe for the elusive FFLO state.
In conventional singlet superfluids, the pairing is spin-isotropic and the COM momentum $\vb{Q}$ vanishes.
In our system, however, the emergence of triplet pairing strictly requires a finite $\vb{Q}$, which is accompanied by a strong spin-momentum locking induced by the altermagnetic term.
Consequently, the spin-resolved momentum distributions become strongly anisotropic and displaced along the diagonal directions.
This feature provides a direct signal to distinguish the FFLO phase from the BCS phase.

\begin{figure}[t]
	\centering
	\includegraphics[width=0.48\textwidth]{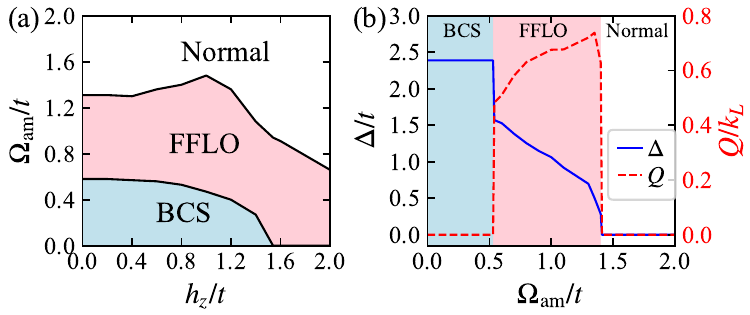}
	\caption{(a) Phase diagram with respect to $h_z$ and $\Omega_{\rm am}$.
	(b) The order parameters $\Delta$ (blue-solid line) and $Q$ (red-dashed line) as functions of $\Omega_{\rm am}$ at $h_z=0.8t$.
	We set $U=6.0t$ and $\mu = 1.0t$.}
	\label{fig:phase_diagram_with_zeeman}
\end{figure}

The phase diagram in the $h_z$-$\Omega_{\rm am}$ plane is displayed in Fig. \ref{fig:phase_diagram_with_zeeman}(a).
We find that a finite Zeeman field $h_z$ favors the FFLO phase, expanding its phase space compared to the spin-balanced case.
Specifically, when $h_z$ exceeds a threshold, the BCS phase region vanishes, while the FFLO phase remains present.
To further illustrate the phase transitions, we show the evolution of the order parameters $\Delta$ and $Q$ as functions of $\Omega_{\rm am}$ for a fixed $h_z$ in Fig. \ref{fig:phase_diagram_with_zeeman}(b).
One can see that the order parameters $\Delta$ and $Q$ exhibit discontinuous changes similar to those shown in Fig. \ref{fig:phase_diagram_without_zeeman}(b), indicating the first-order phase transitions.

\section{Experimental realization}\label{sec:experiment-realization}

The model Hamiltonian considered here can be realized with ultracold $^{40}$K atoms loaded into a two-dimensional square optical lattice.
The two pseudo-spin states $\ket{\uparrow}$ and $\ket{\downarrow}$ can be encoded in two hyperfine levels $\ket{F,m_F}=\ket{9/2,-9/2}$ and $\ket{9/2,-7/2}$,
whose scattering properties allow the attractive interaction $U$ to be tuned via a broad Feshbach resonance.

According to Eq.\eqref{eq:h-at-in-site-space}, the altermagnetic term indeed describes next-nearest-neighbor spin-flip hopping with a relative $\pi$ phase between the $x\pm y$ directions.
This term can be synthesized in a manner analogous to the engineering of two-dimensional spin-orbit coupling \cite{Wu2016OctS,Zhai2015FebRoPiP}.
In particular, one applies an optical coupling between the two pseudo-spins with a spatial mode profile $\Omega_{\rm am}(\vb{r}) \propto \sin(k_L x)\sin(k_L y)$.
The odd parity of this spatial mode with respect to either the $x$ or the $y$ axis eliminates not only the on-site coupling but also the spin-flip terms between nearest-neighbor sites. As a result, the spin-flip processes between next-nearest-neighbor sites dominate.
We note that this engineering induces an additional $\pi$ phase in the bare hopping of one spin species (e.g., $\ket{\downarrow}$ in Eq.\eqref{eq:h-0-in-site-space}).
Such an extra phase can be compensated by a spin-dependent hopping, which can be implemented using laser-assisted tunneling \cite{Aidelsburger2013OctPRL,Miyake2013OctPRL}.

Combining these ingredients, all terms of the Hamiltonian in this paper are accessible with current experimental techniques.

\section{Conclusion}\label{sec:conclusion}

In summary, we have investigated the superfluid phases of the 2D Fermi gas subject to an altermagnetic term coupling to the in-plane spin component $\hat{S}_x$.
We have shown that the combined $C_{4z}\mathcal{T}$ symmetry generates anisotropic Fermi surfaces for the two pseudo-spin species, which favors the FFLO superfluids.
We further show that the polarization drives a coexistence and competition between triplet and singlet pairing channels, giving rise to a momentum-selective pairing pattern.
The model is readily realizable in ultracold atoms, offering an accessible route for quantum emulation of unconventional superfluid phases.

\section*{ACKNOWLEDGMENTS}
We thank Tao Zhou and Ming Gong for helpful discussions.
This work was supported by National Key Research and Development Program of China (Grant No. 2022YFA1405304),
and the Guangdong Basic and Applied Basic Research Foundation (Grant No. 2026A1515012354).

\section*{DATA AVAILABILITY}
The data that supports the findings of this study were generated using theoretical calculations described in the paper. The data are available upon reasonable request from the authors.

\appendix

\section{Pairing modes}\label{sec:pairing-modes}

In the BdG approach, we consider the following pairing Hamiltonian which is extracted from Eq.\eqref{eq:h-pairing-4Q},
\begin{equation}
	H_{\rm int} = \sum_{\vb{k}} (\Delta^\dag \psi_{\frac{\vb{Q}}{2}-\vb{k},\downarrow} \psi_{\frac{\vb{Q}}{2}+\vb{k},\uparrow} +H.c.) + \frac{|\Delta|^2}{U} \,.\label{app:eq:Hamiltonian-int-to-pair}
\end{equation}
We submit the operator transformation \eqref{eq:operator-transfer} into Eq.\eqref{app:eq:Hamiltonian-int-to-pair}, and obtain
\begin{align}
	H_{\rm int} &= \sum_{\vb{k},\vb{Q}} \Big\{
		\Delta^\dag[A_0(\vb{k}) \psi_{\frac{\vb{Q}}{2}-\vb{k},+} \psi_{\frac{\vb{Q}}{2}+\vb{k},+} \notag\\
		&-A_0(-\vb{k}) \psi_{\frac{\vb{Q}}{2}-\vb{k},-} \psi_{\frac{\vb{Q}}{2}+\vb{k},-}
		+A_1(\vb{k}) \psi_{\frac{\vb{Q}}{2}-\vb{k},-} \psi_{\frac{\vb{Q}}{2}+\vb{k},+} \notag\\
		&-A_2(\vb{k}) \psi_{\frac{\vb{Q}}{2}-\vb{k},+} \psi_{\frac{\vb{Q}}{2}+\vb{k},-}]
		+ H.c.
	\Big\} + \frac{|\Delta|^2}{U} \,,\label{app:eq:Hamiltonian-int-to-pair_v2}
\end{align}
where
\begin{equation}
	\begin{cases}
		A_0(\vb{k}) = \sin\frac{\theta_{\vb{Q}/2-\vb{k}}}{2}\cos\frac{\theta_{\vb{Q}/2+\vb{k}}}{2} \\
		A_1(\vb{k}) = \cos\frac{\theta_{\vb{Q}/2-\vb{k}}}{2}\cos\frac{\theta_{\vb{Q}/2+\vb{k}}}{2} \\
		A_2(\vb{k}) = \sin\frac{\theta_{\vb{Q}/2-\vb{k}}}{2}\sin\frac{\theta_{\vb{Q}/2+\vb{k}}}{2}
	\end{cases} \,. \label{app:eq:A-functions}
\end{equation}
Here $\theta_{\vb{k}}$ has been defined in Eq.\eqref{eq:operator-transfer}.
We note that $A_{0,1,2}$ in Eq.\eqref{app:eq:A-functions} depends on $\vb{Q}$, and for notational brevity, we suppress the $\vb{Q}$ dependence in the following.
Hamiltonian \eqref{app:eq:Hamiltonian-int-to-pair_v2} can be further reorganized as
\begin{align}
	H_{\rm int} &= \sum_{\vb{k}} \Big\{
		\Delta^\dag [\{B_1(\vb{k})+B_1'(\vb{k})\}(\psi_{\frac{\vb{Q}}{2}-\vb{k},+} \psi_{\frac{\vb{Q}}{2}+\vb{k},+} \notag\\
		&+ \{B_1(\vb{k})-B_1'(\vb{k})\}\psi_{\frac{\vb{Q}}{2}-\vb{k},-} \psi_{\frac{\vb{Q}}{2}+\vb{k},-}) \notag\\
		&+B_2(\vb{k}) \psi_{\frac{\vb{Q}}{2}-\vb{k},-} \psi_{\frac{\vb{Q}}{2}+\vb{k},+}] + H.c.
	\Big\} + \frac{|\Delta|^2}{U} \label{app:eq:Hamiltonian-int-to-pair_v3}\\
	=& \sum_{\vb{k}} \Big\{
		\Delta^\dag [B_1(\vb{k})\sum_{\nu=\pm}\psi_{\frac{\vb{Q}}{2}-\vb{k},\nu} \psi_{\frac{\vb{Q}}{2}+\vb{k},\nu} \notag\\
	&+B_2(\vb{k}) \psi_{\frac{\vb{Q}}{2}-\vb{k},-} \psi_{\frac{\vb{Q}}{2}+\vb{k},+}] + H.c.
	\Big\} + \frac{|\Delta|^2}{U} \,,\label{app:eq:Hamiltonian-int-to-pair_v4}
\end{align}
where the pairing mode functions are given as
\begin{equation}
	\begin{cases}
		B_1(\vb{k}) = \frac{A_0(\vb{k})-A_0(-\vb{k})}{2} \\
		B_1'(\vb{k}) = \frac{A_0(\vb{k})+A_0(-\vb{k})}{2} \\
		B_2(\vb{k}) = A_1(\vb{k}) + A_2(\vb{k})
	\end{cases} \,. \label{app:eq:B-functions}
\end{equation}
Based on the definition in Eq. \eqref{app:eq:B-functions}, $B_1(\vb{k})$ has odd parity in $\vb{k}$,
whereas $B_1'(\vb{k})$ and $B_2(\vb{k})$ both have even parity.
As the results, the $B_1'(\vb{k})$ terms in Eq.\eqref{app:eq:Hamiltonian-int-to-pair_v3} are discarded owing to the anti-commutation relation of the $\psi$ operators.

We note that for $\vb{Q}=0$, we have $A_0(-\vb{k})=A_0(\vb{k})$ (because of $\theta_{\vb{k}}=\theta_{-\vb{k}}$),
$A_1(\vb{k})=\cos^2\frac{\theta_{\vb{k}}}{2}$, and $A_2(\vb{k})=\sin^2\frac{\theta_{\vb{k}}}{2}$.
It follows that $B_1(\vb{k})=0$, $B_{2}(\vb{k})=1$ at $\vb{Q}=0$.
On the other hand, when $h_z=0$, we have $\sin\frac{\theta_{\vb{k}}}{2}=\cos\frac{\theta_{\vb{k}}}{2}=\frac{1}{\sqrt{2}}$.
It yields $A_{0,1,2}(\vb{k})=\frac{1}{2}$ and consequently $B_1(\vb{k})=0$, $B_{2}(\vb{k})=1$ at $h_z=0$.

Upon summing over all possible COM momenta, we can obtain the Hamiltonian \eqref{eq:h-int-4Q-in-zeeman-case} from Eq.\eqref{app:eq:Hamiltonian-int-to-pair_v4}.

\vfill
\bibliography{ref}
\end{document}